# Structural and elastic properties of amorphous carbon from simulated quenching at low rates

Richard Jana[1,2], Daniele Savio[1,3], Volker L. Deringer[4], Lars Pastewka[1,3,5,6,*]

[1] Institute for Applied Materials, Karlsruhe Institute of Technology, Engelbert-Arnold-Straße 4, 76131 Karlsruhe, Germany

[2] Department of Microsystems Engineering, University of Freiburg, Georges-Köhler-Allee 103, 79110 Freiburg, Germany

[3] Fraunhofer IWM, MicroTribology Center µTC, Wöhlerstraße 11, 79108 Freiburg, Germany

[4] Department of Engineering, University of Cambridge, Trumpington Street, Cambridge CB2 1PZ, UK

[5] Freiburg Materials Research Center, University of Freiburg, Stefan-Meier-Straße 21, 79104 Freiburg, Germany

[6] Cluster of Excellence *liv*MatS @ FIT – Freiburg Center for Interactive Materials and Bioinspired Technologies, University of Freiburg, Georges-Köhler-Allee 105, 79110 Freiburg, Germany

* Corresponding author: lars.pastewka@imtek.uni-freiburg.de

**Abstract**

We generate representative structural models of amorphous carbon (a-C) from constant-volume quenching from the liquid with subsequent relaxation of internal stresses in molecular dynamics simulations using empirical and machine-learning interatomic potentials. By varying volume and quench rate we generate structures with a range of density and amorphous morphologies. We find that all a-C samples show a universal relationship between hybridization, bulk modulus and density despite having distinct cohesive energies. Differences in cohesive energy are traced back to slight changes in the distribution of bond-angles that will likely affect thermal stability of these structures.

1. Introduction

There has been a growing scientific and industrial interest in coatings composed of amorphous carbon (a-C). Many studies, extensively reviewed for example in Refs.[1] and [2] report on their outstanding mechanical properties, hardness, and resilience, which provide low friction and wear in poorly lubricated and highly loaded contacts. These characteristics are, however, strongly affected by the coating process and the resulting chemical composition and atomic structure. For amorphous carbon materials with low concentration of hydrogen, links exist between the density, elastic moduli, hardness and atomic structure[3–5]. For instance, a-C films made up mainly by $sp^2$ (3-fold coordinated) carbons are relatively soft and easily worn[4]. Conversely, tetrahedral amorphous forms of carbon ("ta-C") with high proportions of $sp^3$ (fourfold coordinated) carbons present high density and elastic modulus, and are extremely hard[6]. Usual experimental measurements of these quantities involve spectroscopy[7,8] for the $sp^3$ content, or nano-indentation[4,9], surface Brillouin scattering[10,11], and acoustic wave propagation[12] for the mechanical properties. These complex techniques cannot be applied directly in-situ to characterize the behavior of a-C in highly loaded contacts.

Atomic-scale simulations based on Molecular Dynamics (MD) have thus been increasingly used to gain insights into mechanical and tribological properties of these materials. Several studies report atomic-scale structural modifications occurring under complex tribological loading like amorphization, re-hybridisation[13–15] and fracture[16]. Yet, atomic-scale simulations have rarely been used to quantify the basic structural and mechanical properties of a-C[17–19]. Moreover, the few comparisons with experiments reveal severe limitations of some interatomic potential formulations in capturing the characteristic features of amorphous carbon[20]. Despite recent interest in the material, there is also only a limited number of studies using ab initio methods like DFT[21–23], due to their large computational cost.

Among the most important quality criteria for modeling a-C materials is what fraction of atoms are predicted to be $sp^3$ bonded. In an extensive study, Ref.[15] reviewed six interatomic potentials (Tersoff[24], REBO-II[25], EDIP[26], LCBOP-I[27], ReaxFF[28] and COMB3[29]) commonly used for a-C in MD simulations, but found $sp^3$ fractions at least 10% lower than in experimental studies for all potentials. Some of the potentials studied formed questionable morphologies at lower densities, such as the lack of hexagons that are expected to be abundant at low densities or, conversely, the abundance of triangles which are not expected to occur because are energetically unfavorable.

The present paper aims at providing a further reference study on this issue, with a special focus on a comparison between experiments and atomistic simulations, going to much slower quench rates (giving the amorphous structures more time to find a low energy configuration), using more realistic potentials and including elastic properties of the formed solid phases. A uniform simulation protocol will be employed to generate a wide variety of amorphous carbon samples with different computational methods. The density, structural and mechanical properties of the computer-generated structures will be analyzed and compared to existing experimental and DFT values. Finally, we will suggest extracting additional structural parameters from Molecular

Dynamics simulations, which may play a significant role on the stability and structural transformations of amorphous carbon under thermal and mechanical stress.

## 2. Methods

We use a liquid quench protocol to generate a-C samples with a range of structures. Rapid quench of a melt in a periodic simulation cell is a standard procedure to create glassy morphologies in atomistic calculations[15]. This method is a surrogate for the experimental growth of thin a-C films with physical vapor deposition or ion irradiation techniques, which have also been studied by atomic-scale simulations but incur even much larger computational cost[13,30–32]. Through different starting densites and quench rates varying over two orders of magnitude, our study aims at creating representative volume elements of varying atomic morphologies and therefore (potentially) varying properties. This will allow us to identify structure-property relationships.

We compare three material models. As a model of intermediate accuracy and transferability, we use a Gaussian Approximation Potential (GAP)[33] model parameterized for a-C[34], which has been previously shown to provide accuracy very close to DFT calculations but is computationally less expensive and scales linearly with system size. In order to study the influence of system size and probe a larger parameter space, in particular low quench rates, we also carry out MD calculations with empirical bond-order potentials[16]. We use the Tersoff III potential[35] with additional screening functions (Tersoff+S, Ref.[20]) that were introduced for reproducing brittle fracture mechanics in diamond[36]. This formulation was chosen because it reproduces the density-$sp^3$ relationship found in experiments and DFT calculations well while having moderate computational cost. For high-fidelity calculations, we carry out calculations using spin-paired density functional theory (DFT)[37,38] within the local density approximation[39], a local double-zeta basis set and Goedecker-Teter-Hutter pseudopotentials[40]. We use an energy cutoff of 250

Ry for the systems consisting of pure carbon. Those calculations are limited to rather small simulation cells. We subsequently relax and calculate the mechanical properties with the GAP potential.

We perform several simulations, each generating a different a-C sample. The carbon atoms are placed randomly in a cubic box, with initial density $\rho_{init}$ between 1.7 g/cm³ and 3.5 g/cm³. The system is held for 100 ps at a temperature of 10000 K to equilibrate the molten state. Then, we quench the system down to 0 K at constant volume. Cooling rates are varied between $\dot{T}_c$ = 10 K ps⁻¹, 100 K ps⁻¹ and 1000 K ps⁻¹ for Tersoff+S, only the fast and medium cooling rate is used for the GAP calculations, only the fastest for the DFT calculations. The fastest one is on the order of the cooling rates used in most DFT-based simulations[21]. All these steps are performed with a time-step of 0.2 fs for Tersoff+S and GAP simulations, and 1 fs for DFT. Finally, we relax the supercell to zero pressure using a minimization algorithm. Full triclinic cell deformations are allowed to obtain stress-free structures. We note that in many previous calculations[15,16,20,21,36], the intrinsic stress in the structures was not relaxed. DFT calculation are carried out using boxes with 216 carbon atoms. Calculations run with the classical potential contain 4087 carbon atoms.

## 3. Results and discussion

We analyze the a-C structures generated by the aforementioned protocol with respect to their density, sp³ content and elastic constants. The final density of the structures differs from $\rho_{init}$ because the cell relaxation can change the cell volume. The densities of the final a-C samples range from $\rho$ = 1.5 g/cm³ to 3.21 g/cm³; example snapshots of these structures are shown in Fig. 1a and b. The upper bound for $\rho$ is in agreement with the maximum density of 3.3 g/cm³ reported in experiments[3], DFT simulations[21,22], and theoretical calculations for an amorphous structure with maximum sp³ content[17].

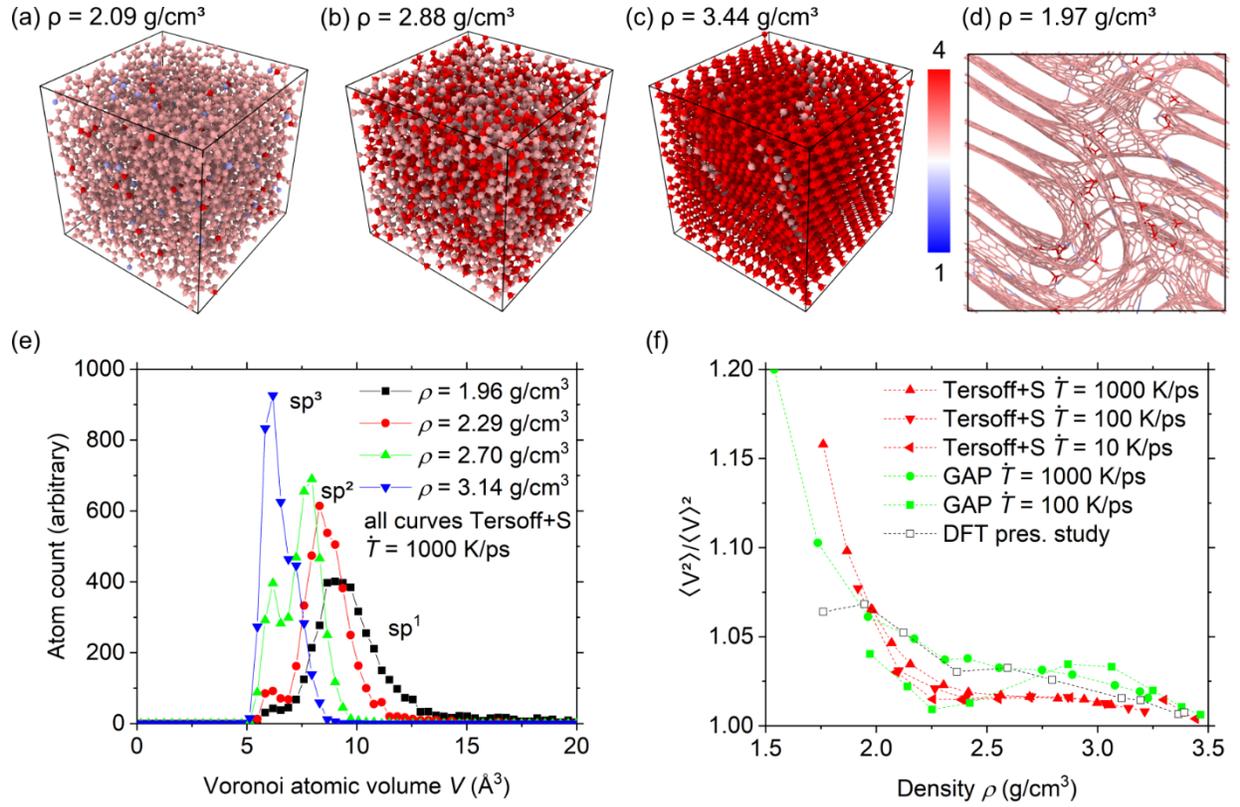

Fig. 1: Snapshots of amorphous ($\rho$ = 2.09 g/cm³ and 2.88 g/cm³) (a and b) and nano-crystalline ($\rho$ = 3.44 g/cm³) (c) carbon samples generated through liquid quenches using the Tersoff+S potential at a rate of $\dot{T}_c$ = 10 K ps$^{-1}$. The coordination number is encoded in the color. (d) Snapshot of a structure with $\rho$ = 1.97 g/cm³ quenched with the GAP potential at $\dot{T}_c$ = 100 K ps$^{-1}$. Here we find extended graphitic sheets. (e) Distribution of atoms by Voronoi volume $V$ in structures with different densities, generated with Tersoff+S, quenched at $\dot{T}_c$ = 1000 K ps$^{-1}$. (f) Width of the distribution of Voronoi volumes characterized by $\langle V^2 \rangle / \langle V \rangle^2$, where $\langle \, \rangle$ is the average. Large values of $\langle V^2 \rangle / \langle V \rangle^2$ indicate the presence of nanovoids. Lines are intended as a guideline for the eye.

Note that the structures obtained at slower quench rates ($\dot{T}_c$ = 100 K ps$^{-1}$ and $\dot{T}_c$ = 10 K ps$^{-1}$) vary significantly between the Tersoff+S and GAP. We find structures above 3.21 g/cm³ density in

Tersoff+S, but those have partially recrystallized to form a nano-crystalline morphology (Fig. 1c). An upper bound to the structures generated by the quench protocol is the density of diamond, 3.515 g/cm³[41]. We observe recrystallization for samples generated from high initial density $\rho_{init}$ > 3.0 g/cm³ and lowest $\dot{T}_c$ = 10 K ps⁻¹, where the relatively slow cooling rate allows for a reorganization into nanocrystals. Conversely, using GAP we observe the formation of extended graphitic flakes in the GAP calculations with $\dot{T}_c$ = 100 K ps⁻¹ and low densities of 2.55 g/cm³ and below. We show an example snapshot with graphitic flakes in Fig. 1d. The latter may be compared with the formation of "graphitized" carbon nanostructures formed in long annealing (rather than quenching) simulations using the EDIP[42] and GAP[43] models.

To analyze the microscopic morphology of the final structures using statistical measures, we carry out a Voronoi analysis (Fig. 1e). Voronoi volume distributions present two peaks at $V$ = 6.2 Å³ and $V$ = 8.2 Å³, corresponding to the mean atomic volumes of sp³ and sp² atoms, respectively. The values for diamond (pure sp³) and graphite (pure sp²) from a relaxed GAP calculation are $V$ = 5.467 Å³ and $V$ = 8.708 Å³, respectively. The magnitude of the sp² peak increases compared to the sp³ one as the density of the structure is reduced. Moreover, a large tail at high Voronoi volumes appears for samples with very low density. This can be attributed to the formation of nanovoids within the amorphous network, i.e. pores where small gas molecules could fit[23]. To quantify the extent to which large volumes contribute to the overall Voronoi volume distribution, we compute $\langle V^2 \rangle / \langle V \rangle^2$ where $V$ is the Voronoi volume and $\langle \cdot \rangle$ the average over all atoms. It is a semi-quantitative measure for the number of atoms with a large Voronoi volume. Figure 1f shows this measure as a function of density. The contribution of large Vonoroi volumes increases rapidly for densities below 2.3 g/cm³, which is close to the density of graphite. We note that this lower density bound of amorphous structures is also in accordance with reported experimental values of 2.0 - 2.2 g/cm³[2,3]. Below this limit, nanovoids or graphitic sheets appear in structures, which are detected by the presence of large Voronoi volumes for some atoms in our calculations.

Next, we compute the fraction of sp³ carbon by counting nearest neighbors for each atom. The distance cutoff is chosen at 1.85 Å, where the radial distribution function for the chosen interatomic potential drops to zero (Fig. 2a). Figure 2b shows that the sp³ fraction of the structures increases with density. This trend is in good qualitative agreement with experimental[3] and DFT data[21,22] for both potentials. Fast quenches using the GAP potential fit the experimental reference data better than the Tersoff+S data, which deviates from the experimental references at low density. Both GAP and Tersoff+S show a similar sp³ fraction at low quench rates, although the structures differ significantly between graphitic sheets in GAP and a disordered sp² network in Tersoff+S.

The differences in the structures between the potentials can be seen in Fig. 2 c and d, where we show ring statistics. Tersoff+S and GAP structures differ in the number of 5- and 6-rings, which are more common with GAP, especially at the low quench rate. The GAP and DFT quenches at 1,000 K ps⁻¹ lead to essentially indistinguishable ring-size distributions, especially when taking a certain statistic scatter into account; this evidences the ability of the GAP to reproduce the DFT potential-energy surface. Both the diamond crystals and the graphitic sheets consist of 6-membered rings, which explains the high fraction at high density for Tersoff+S $\dot{T}_c$ = 10 K ps⁻¹ at high density and GAP $\dot{T}_c$ = 100 K ps⁻¹ at low densities, respectively.

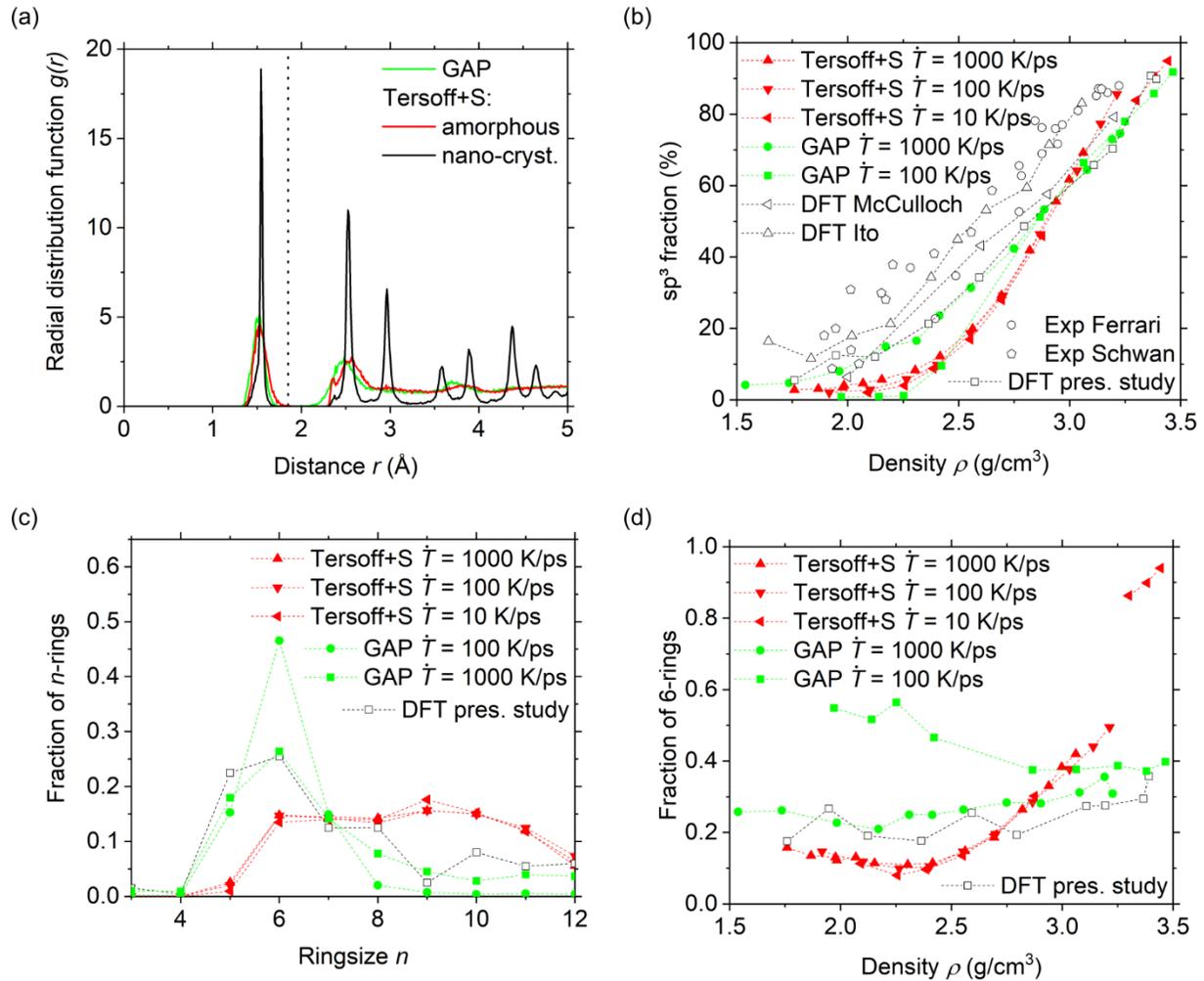

Fig. 2: (a) Radial distribution functions for amorphous and nano-crystalline structures generated with DFT, GAP and the Tersoff+S. The dotted line indicates the neighbor cutoff at 1.85 Å. (b) $sp^3$ fraction as a function of density. Experimental data is from Refs.[3,44], and DFT results from Refs.[21,22]. Lines are intended as a guideline for the eye. (c) Ring statistics for all our calculations at densities $\rho$ of ~2.55 g/cm$^3$. (d) Fraction of 6-membered rings in the structures at different densities for all our calculations.

We now turn to the elastic properties of a-C. We perform small triclinic deformations of the samples using a strain between $\Delta\varepsilon = 0.0001$ % and $0.1$ % while optimizing the atomic positions to the respective ground state. In the range tested, the magnitude of the strain increment does not impact the elastic constants obtained. The elastic constants are calculated through linear interpolation of the stress-strain curves, which have a constant slope over the applied strain range. The elastic constants generally fulfill the relationships for isotropic bodies, $C_{11} = C_{22} = C_{33}$, $C_{44} = C_{55} = C_{66}$, $C_{12} = C_{13} = C_{23}$. All other elastic constants vanish. The Zener anisotropy factor $A = 2C_{44}/(C_{11} - C_{12})$ lies in the range $1\pm0.1$ for all amorphous structures. Figure 3a shows a linear increase of the bulk modulus $K$ with density $\rho$, where Tersoff+S and GAP values agree very well. DFT results from[22] are similar, but have a different slope $K(\rho)$ compared with our data.

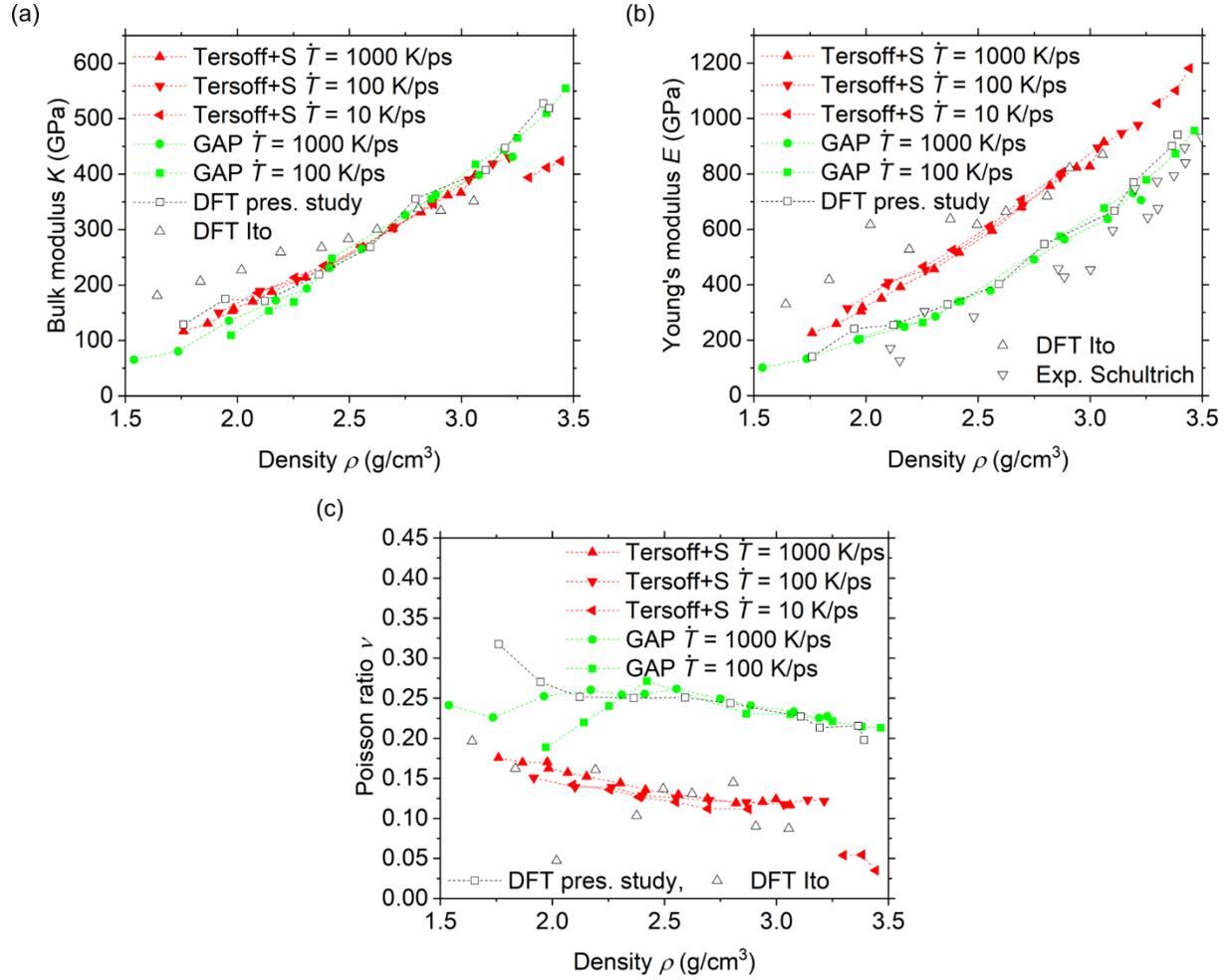

Fig. 3: Elastic properties as a function of density. (a) Bulk modulus $K$. (b) Poisson's ratio $\nu$. (c) Young's modulus $E$. Experimental data is from[5], and DFT results from[22]. DFT results from present study are elastic constants computed using GAP on the structures quenched with DFT. Lines are intended as a guideline for the eye.

The Young's modulus $E$ (shown in Fig. 3b), however, differs between Tersoff+S and GAP structures. While the data for Tersoff+S are closer to previous DFT data (Ref.[22]), the GAP data fits better to the experimental data[5]. The discrepancy in values for the Young's modulus can be attributed to various factors which influence experimental values, such as the thickness of the a-C films, the presence of hydrogen, or the fact that the Young's modulus cannot be measured directly.

Obtaining $E$ from nano-indentation[45] or wave propagation experiments[5] requires knowledge of the Poisson's ratio $\nu$. This is hard to measure experimentally, but it is generally assumed that $\nu$ does not change with density. Values found in literature vary from 0.12[46], over ~0.20[47,48] to 0.25[49,50]. Figure 3c shows our calculation of $\nu$. The Poisson ratio is constant at around 0.25 for GAP but decreases from 0.17 to 0.11 over the density range $\rho$ = 1.9 - 3.5 g/cm³ for the Tersoff+S structures. The results obtained with Tersoff+S agree with DFT results from Ref.[22], where the calculated values are nonetheless a lot more scattered, probably due to the smaller system size. We note that the discrepancy in $E$ between GAP and Tersoff+S can be traced back solely to this discrepancy in $\nu$.

These results indicate that our simulations with the chosen sample generation protocol and carbon-carbon interaction potential can create amorphous structures with realistic density, structural properties and elastic constants. However, for our glassy samples, none of these properties depend on quench rate, given that the final structure is disordered and does not crystallize into diamond or graphite. We now look for additional structural indicators that are able to discriminate between the glassy structures obtained at different quench rates.

First, we determine the bond lengths inside the MD-generated structures as shown in Fig. 4 for the GAP and Tersoff+S structures. The average distance between sp³-sp³ atoms is approximately $l_{sp^3-sp^3}$ = 1.61 Å for low density samples, while sp²-sp² bonds are shorter ($l_{sp^2-sp^2}$ = 1.47 Å) for Tersoff+S. The average bond length between sp²-sp³ atoms lies in the middle between the aforementioned values. GAP predicts shorter bond-lengths in all cases in line with the smaller lattice constant for the diamond structure (the bond length in ideal diamond being 1.525 Å for GAP vs. 1.544 Å for Tersoff+S). All bond lengths decrease by approximately 2 % with increasing density. However, the sample cooling rate $\dot{T}_c$ does not impact significantly the interatomic distances within the MD-generated structures. The bond lengths for the DFT structures are not shown in Fig. 4, but fall on top of the GAP curves.

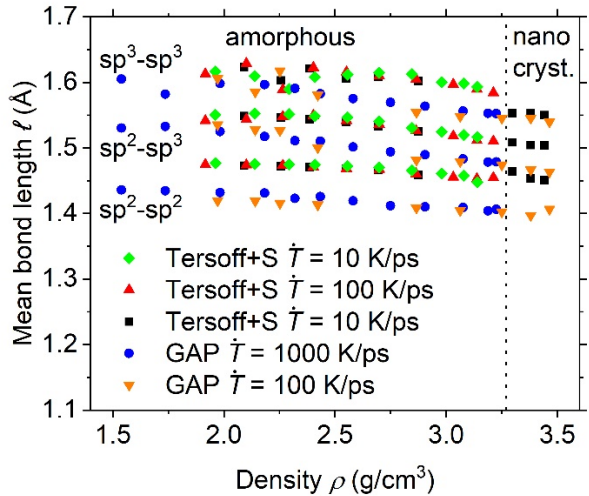

Fig. 4: Bond length as a function of density.

We next focus on the cohesive energy per atom $E_{pot/atom}$, measured relative to the per-atom energy of diamond for the respective potential. This quantity is shown in Fig. 5 as a function of density. The Tersoff+S amorphous structures have a minimum at a density around $\rho = 2.25$ g/cm$^3$, close to the density of graphite. The corresponding structures are almost exclusively composed of sp$^2$ carbon, but disordered (glassy) without the formation of extended graphitic sheets. The cohesive energy then rises as the sp$^3$ fraction increases with density. $E_{pot/atom}$ increases also for $\rho < 2.25$ g/cm$^3$, as nanovoids with corresponding free surfaces appear inside the structures (Figure 1). The Tersoff+S energetically stabilizes "graphitic" sp$^2$ even if it appears in disordered structured rather than in sheets.

The relative cohesive energy for the nano-crystalline samples is much lower and approaches the value reported for diamond[51]. Figure 5 shows also a significant difference in relative cohesive energy for samples generated at different quench rates. Relatively slow cooling from the molten state at $\dot{T}_c = 10$ K ps$^{-1}$ allows carbon atoms to organize into stable, energetically favorable structures. On the other hand, fast cooling at 100 K ps$^{-1}$ and 1000 K ps$^{-1}$ freezes the disordered structure of the liquid phase with little reorganization of the carbon atoms. The resulting relative cohesive energy is higher by 0.1 - 0.15 eV per atom. This value is of order of the energy barrier for the reorganization of sp$^3$ into sp$^2$ phases reported in the literature for high temperature deposition and post deposition annealing[52]. The cohesive energy per atom is therefore the first measure presented here that does depend significantly on quench rate.

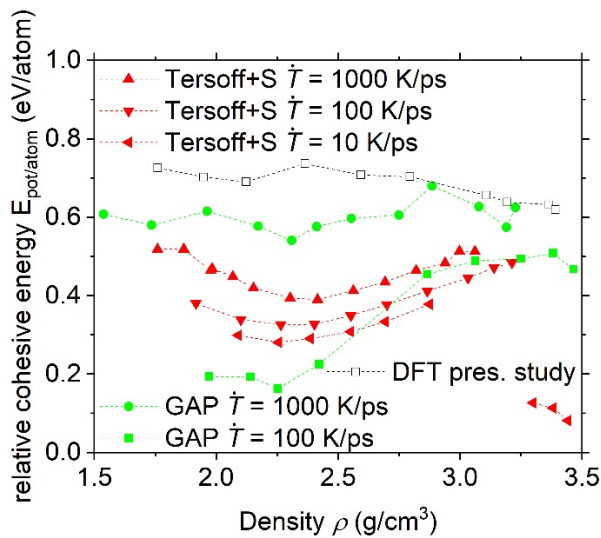

Fig. 5: Relative cohesive energy (relative to the energy of diamond) as a function of density. Lines are intended as a guideline for the eye.

The GAP structures quenched with 1000 K ps$^{-1}$ have energies differences (with respect to crystalline diamond) higher than the Tersoff+S structures, and the minimum at around $\rho = 2.3$ g/cm³ is less pronounced. As we show in Fig. 2a, the sp³ content in the GAP structures at this density is much higher than that in the Tersoff+S structures, about 16% compared to 8%, which means that the sp² content is lower. This could suppress the minimum in the $E_{\text{pot/atom}}(\rho)$ curves. The energy of the GAP structures quenched with 100 K ps⁻¹ at densities of 2.8 g/cm³ (glassy disordered networks) and above is comparable to the energy of the Tersoff+S structures and lower than the respective structures obtained with GAP at higher quench rates. At lower density, where we find graphitic sheets, the energy is much lower. There also is a minimum at approx. $\rho = 2.25$ g/cm³ but we observe graphitic sheets at this density.

The cohesive energy may therefore play a crucial role for the stability, and consequently the response under mechanical loading, of our amorphous carbon structures. It should also be noted again that traditional parameters for the characterization of a-C fail at quantifying this important feature of our structures. Figures 1-4 show in fact that both the density, sp³ fraction, elastic properties and bond lengths are independent of the sample quench rate. Hence, an additional structural parameter which correlates well with the observed variations in cohesive energy would be useful. The only structural change that we were able to find in our samples was a slight broadening of the bond angle distribution, as described in the following.

A measure for the disorder of the amorphous carbon with $\dot{T}_c$ is the distribution of bond angles at sp³ and sp² sites. The underlying principle relies on the deformation of the crystalline structures of diamond and graphite occurring in a-C. In diamond, the nearest neighbors of each sp³ atom create a perfect tetrahedron with angles of 109.28°[41]. In graphite, sp² carbons form planar hexagonal structures with angles of 120°. In the case of a-C, we find that the angle distributions for sp³ and sp² atoms are approximately Gaussian and centered around these angles (Fig. 6a). Their means do not vary significantly with the sample cooling rate. For the GAP structures both peaks are lower and wider, compared to Tersoff+S (Fig. 6b).

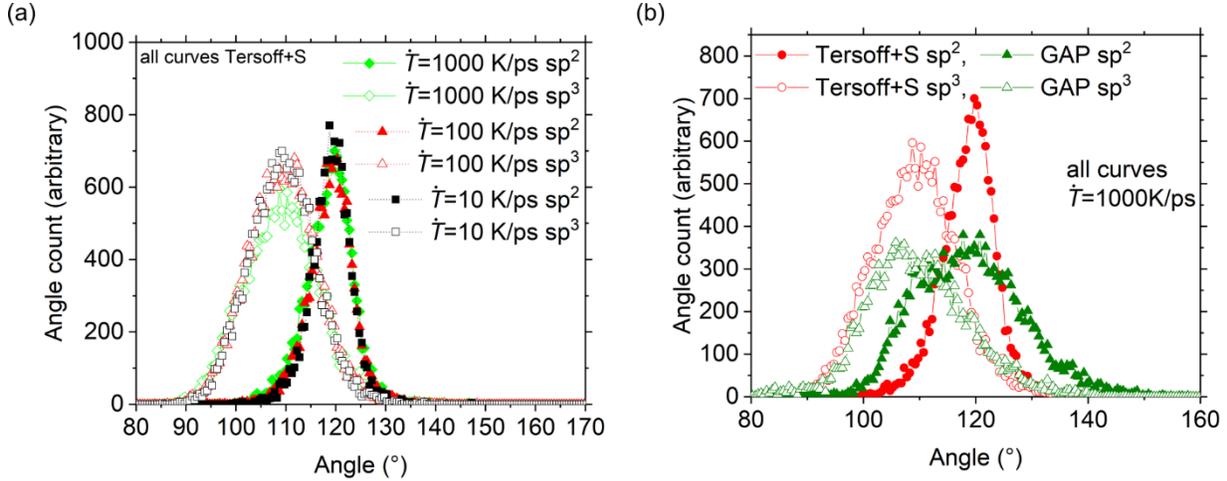

Fig. 6: (a) Angle distribution for sp² and sp³ atoms. Three samples are considered, with same density $\rho = 2.87$ g/cm³, sp³ fraction (46%) and elastic constants, but generated using different cooling rates. (b) Angle distribution for sp² and sp³ atoms for two structures, generated with Tersoff+S and GAP. Tersoff+S data $\rho = 2.87$ g/cm³, GAP data for density $\rho = 2.55$ g/cm³ (sp³ fraction about 31%). The densities are chosen such that the peaks for sp² and sp³ have comparable height.

However, the distributions get broader as $\dot{T}_c$ increases. Essentially, this means that higher deformations of the ideal tetrahedral and hexagonal carbon structures occur when faster quenching is employed in the sample generation protocol. Consequently, the cohesive energy should also be higher. The width of the angle distribution can be quantified from its standard deviation $\sigma$, and we indeed observe a linear correlation between $\sigma$ for the angle distribution and $E_{pot/atom}$ for all our amorphous carbon samples (Fig. 7). This structural parameter can thus provide additional information on the stability of computer-generated amorphous carbon structures, in addition to the usual density, sp³ fraction and elastic properties. The distributions of the GAP structures are broader than those of the Tersoff+S structures. This can be seen from the standard deviations in Fig. 7, which are about 2.5° higher. The data points for the nano-crystalline and the graphitic structures are separated from the amorphous data points. Here, the angles have a lower standard variation. This is explained by the well-defined angles in the nano-crystalline and graphitic regions, which consist of $sp^3$ atoms in a diamond lattice and $sp^2$ atoms in the graphite lattice.

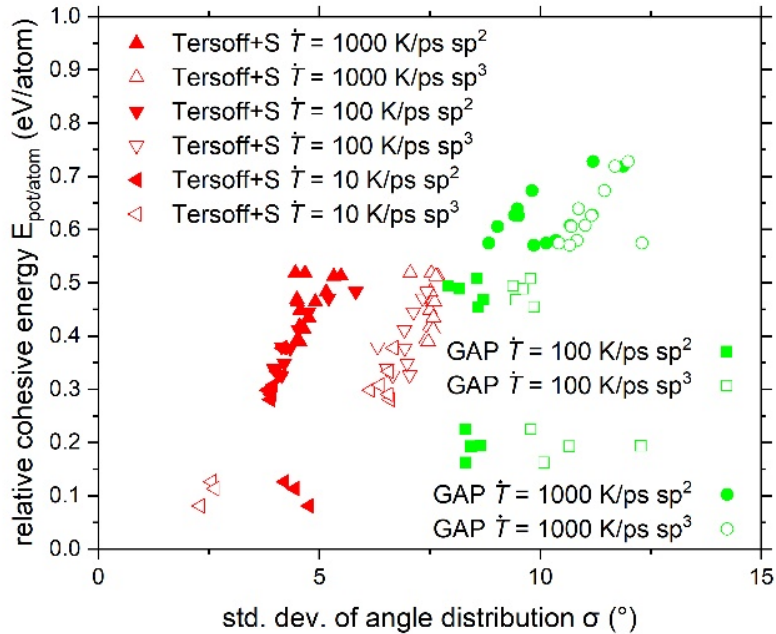

Fig. 7: Cohesive energy as a function of the standard deviation for the angle distribution. The angle distribution is collected independently for $sp^2$ and $sp^3$ atoms.

## 4. Summary and conclusions

This study describes properties of amorphous carbon (a-C) samples obtained through rapid quenches in molecular dynamics simulations, surveying different quench rates and simulation methods. The quenched structures are subsequently relaxed to obtain stress-free reference samples. The relaxed a-C structures are analyzed using established structural parameters. We find densities ranging from 2.0 to 3.3 g/cm³, with nanovoids forming in the structures with density lower than

2.0 g/cm$^3$. The sp$^3$ fraction rises from ~5 to ~80 % over the reported density range. The bulk modulus ranges from ~150 to 430 GPa. Results obtained with different methods (Tersoff+S[20], GAP[34] and DFT) agree qualitatively with one another. We find differences in the tendency of the formation of diamond crystals at high density (only observed for Tersoff+S) and graphite crystals at low density (only observed for GAP). Poisson ratios also differ between models from ~0.15 in Tersoff+S to ~0.25 in GAP, but lie within the range reported experimentally.

The quench rate in the sample generation protocol does not affect the aforementioned quantities, but it impacts strongly the cohesive energy of the a-C samples. Fast cooling leads to less energetically stable structures by causing distortion of the ideal hexagonal and tetrahedral structures formed by neighbors of sp$^2$ and sp$^3$ atoms. This distortion is visible in the standard deviation of the local angle distribution, which correlates linearly with the cohesive energy for all amorphous structures. This is true for both the Tersoff+S and GAP potentials, although the absolute values of the widths of the angle distribution functions differ.

Our results indicate that it is difficult to discriminate a-C structures of different cohesive energy (and hence different structural stability) based on simple geometric parameters alone. The change in bond-angle distribution indicates that less stable structures are more distorted than more stable a-Cs, but since the structural change is small we are not confident that this could be detected in experiments. It is also interesting to note that the two potentials tested here have different tendencies towards crystallization. Tersoff+S forms diamond-like regions at high density and the lowest quench rates. GAP forms graphite at low density and low quench rates. Both potentials create glassy disordered networks otherwise. Since GAP was trained on DFT-LDA data that fortuitously predicts the correct interlayer spacing[53,54] while Tersoff+S does not, we tend to trust the GAP calculations more than Tersoff+S in the low-density region. The functional form of

GAP has additionally a higher flexibility to capture different local environments and is therefore expected to be more accurate, giving the training set contains the relevant structures. However, it is also a factor of 100 slower than Tersoff+S, limiting the attainable simulation times on present day computers. Liquid quenches are a substitute for the creation of amorphous carbon in experiments, that is typically grown in physical vapor deposition experiments[2]. More investigation is certainly needed on films grown in realistic vapor deposition simulations[13,30–32].

**Acknowledgements**

We thank Peter Gumbsch for useful discussion and the Deutsche Forschungsgemeinschaft (DFG) for partially funding this work through grant PA 2023/2. Simulations involving classical potentials were conducted with LAMMPS[55]. DFT calculations were carried out with CP2K[56]. Calculations were carried out on JURECA at the Jülich Supercomputing Center (grant "hfr13") and on NEMO at the University of Freiburg (DFG grant INST 39/963-1 FUGG). Post-processing and visualization was carried out with ASE[57], matscipy[58] and Ovito[59]. V.L.D. acknowledges a Leverhulme Early Career Fellowship and support from the Isaac Newton Trust.

**References**

[1]     Erdemir A and Donnet C 2006 Tribology of diamond-like carbon films: recent progress and future prospects *J. Phys. D: Appl. Phys.* **39** R311--R327

[2]     Robertson J 2002 Diamond-like amorphous carbon *Mater. Sci. Eng. R Rep.* **37** 129–281

[3]     Ferrari A C, Libassi A, Tanner B K, Stolojan V, Yuan J, Brown L M, Rodil S E, Kleinsorge B and Robertson J 2000 Density, sp$^3$ fraction, and cross-sectional structure of amorphous carbon films determined by x-ray reflectivity and electron energy-loss spectroscopy *Phys. Rev. B*

**62** 11089

[4]     Andujar J L, Polo M C, Esteve J, Robertson J and Milne W I 2001 Study of the mechanical properties of tetrahedral amorphous carbon films by nanoindentation and nanowear measurements *Diam. Rel. Mat.* **10** 145–52

[5]     Schneider D and Schultrich B 1998 Elastic modulus: a suitable quantity for characterization of thin films *Surf. Coat. Technol.* **98** 962–70

[6]     Sullivan J P, Friedmann T A and Hjort K 2001 Diamond and amorphous carbon MEMS *MRS Bull.* **26** 309–11

[7]     Ferrari A C and Robertson J 2000 Interpretation of Raman spectra of disordered and amorphous carbon *Phys. Rev. B* **61** 14095–107

[8]     Beghi M G, Ferrari A C, Teo K B K, Robertson J, Bottani C E, Libassi A and Tanner B K 2002 Bonding and mechanical properties of ultrathin diamond-like carbon films *Applied Physics Letters* **81** 3804–6

[9]     Oliver W C and Pharr G M 2011 Nanoindentation in materials research: Past, present, and future *MRS Bull.* **35** 897–907

[10]    Beghi M G, Casari C S, Bassi A L, Bottani C E, Ferrari A C, Robertson J and Milani P 2002 Acoustic phonon propagation and elastic properties of nano-sized carbon films investigated by Brillouin light scattering *Thin Solid Films* **420–421** 300–5

[11]    Beghi M G, Bottani C E, Bassi A L, Ossi P M, Tanner B K, Ferrari A C and Robertson J 2002 Measurement of the elastic constants of nanometer-thick films *Materials Science and Engineering C* **19** 201–4

[12]     Schultrich B, Scheibe H-J, Grandremy G and Schneider D 1994 Elastic Modulus of Amorphous Carbon Films *phys. stat. sol. (a)* **385**

[13]     Fairchild B A, Rubanov S, Lau D W M, Robinson M, Suarez-Martinez I, Marks N A, Greentree A D, McCulloch D G and Prawer S 2012 Mechanism for the Amorphisation of Diamond *Adv. Mater.* **24** 2024–9

[14]     Kunze T, Posselt M, Gemming S, Seifert G, Konicek A R, Carpick R W, Pastewka L and Moseler M 2014 Wear, Plasticity, and Rehybridization in Tetrahedral Amorphous Carbon *Tribol. Lett.* **53** 119–26

[15]     de Tomas C, Suarez-Martinez I and Marks N A 2016 Graphitization of amorphous carbons: A comparative study of interatomic potentials *Carbon* **109** 681–93

[16]     Pastewka L, Mrovec M, Moseler M and Gumbsch P 2012 Bond order potentials for fracture, wear, and plasticity *MRS Bull.* **37** 493–503

[17]     Kelires P C 1994 Elastic properties of amorphous carbon networks *Phys. Rev. Lett.* **73** 2460–3

[18]     Kelires P C 2000 Intrinsic stress and local rigidity in tetrahedral amorphous carbon *Phys. Rev. B* **62** 15686–94

[19]     Remediakis I N, Fyta M G, Mathioudakis C, Kopidakis G and Kelires P C 2007 Structure, elastic properties and strength of amorphous and nanocomposite carbon *Diam. Relat. Mater.* **16** 1835–40

[20]     Pastewka L, Klemenz A, Gumbsch P and Moseler M 2013 Screened empirical bond-order potentials for Si-C *Phys. Rev. B* **87** 205410

[21]     McCulloch D G, McKenzie D R and Goringe C M 2000 Ab initio simulations of the structure of amorphous carbon *Phys. Rev. B* **61** 2349–55

[22]     Ito A M, Takayama A, Oda Y and Nakamura H 2014 The First principle calculation of bulk modulus and Young's modulus for amorphous carbon material *J. Phys. Conf. Ser.* **518** 012011

[23]     Held A and Moseler M 2019 Ab initio thermodynamics study of ambient gases reacting with amorphous carbon *Phys. Rev. B* **99** 054207

[24]     Tersoff J 1988 Empirical interatomic potential for carbon, with applications to amorphous carbon *Phys. Rev. Lett.* **61** 2879

[25]     Brenner D W, Shenderova O A, Harrison J A, Stuart S J, Ni B and Sinnott S B 2002 A second-generation reactive empirical bond order (REBO) potential energy expression for hydrocarbons *J. Phys. Condens. Matter* **14** 783–802

[26]     Marks N A 2000 Generalizing the environment-dependent interaction potential for carbon *Phys. Rev. B* **63** 35401

[27]     Los J H and Fasolino A 2003 Intrinsic long-range bond-order potential for carbon: Performance in Monte Carlo simulations of graphitization *Phys. Rev. B* **68** 024107

[28]     Srinivasan S G, van Duin A C T and Ganesh P 2015 Development of a ReaxFF Potential for Carbon Condensed Phases and Its Application to the Thermal Fragmentation of a Large Fullerene *J. Phys. Chem. A* **119** 571–80

[29]     Liang T, Shan T-R, Cheng Y-T, Devine B D, Noordhoek M, Li Y, Lu Z, Phillpot S R and Sinnott S B 2013 Classical atomistic simulations of surfaces and heterogeneous interfaces with the charge-optimized many body (COMB) potentials *Mater. Sci. Eng. R Rep.* **74** 255–79

[30]     Jäger H U and Albe K 2000 Molecular-dynamics simulations of steady-state growth of ion-deposited tetrahedral amorphous carbon films *J. Appl. Phys.* **88** 1129–35

[31]     Jäger H U and Belov A Y 2003 ta-C deposition simulations: Film properties and time-resolved dynamics of film formation *Phys. Rev. B* **68** 24201

[32]     Caro M A, Deringer V L, Koskinen J, Laurila T and Csányi G 2018 Growth Mechanism and Origin of High p3 Content in Tetrahedral Amorphous Carbon *Phys. Rev. Lett.* **120** 166101

[33]     Bartók A P, Payne M C, Kondor R and Csányi G 2010 Gaussian Approximation Potentials: The Accuracy of Quantum Mechanics, without the Electrons *Phys. Rev. Lett.* **104** 136403

[34]     Deringer V L and Csányi G 2017 Machine learning based interatomic potential for amorphous carbon *Phys. Rev. B* **95** 094203

[35]     Tersoff J 1989 Modeling solid-state chemistry: Interatomic potentials for multicomponent systems *Phys. Rev. B* **39** 5566–8

[36]     Pastewka L, Pou P, Pérez R, Gumbsch P and Moseler M 2008 Describing bond-breaking processes by reactive potentials: Importance of an environment-dependent interaction range *Phys. Rev. B* **78** 161402(R)

[37]     Hohenberg P and Kohn W 1964 Inhomogeneous Electron Gas *Phys. Rev.* **136** B864–71

[38]     Kohn W and Sham L J 1965 Self-Consistent Equations Including Exchange and Correlation Effects *Phys. Rev.* **140** A1133–8

[39]     Ceperley D M and Alder B J 1980 Ground State of the Electron Gas by a Stochastic

Method *Phys. Rev. Lett.* **45** 566–9

[40]    Goedecker S, Teter M and Hutter J 1996 Separable dual-space Gaussian pseudopotentials *Phys. Rev. B* **54** 1703–10

[41]    Field J E 1992 *The Properties of Natural and Synthetic Diamond* (London: Academic Press)

[42]    Powles R C 2009 Self-assembly of sp 2 -bonded carbon nanostructures from amorphous precursors *Phys. Rev. B* **79** 075430

[43]    Deringer V L, Merlet C, Hu Y, Lee T H, Kattirtzi J A, Pecher O, Csányi G, Elliott S R and Grey C P 2018 Towards an atomistic understanding of disordered carbon electrode materials *Chem. Commun.* **54** 5988–91

[44]    Schwan J, Ulrich S, Roth H, Ehrhardt H, Silva S R P, Robertson J, Samlenski R and Brenn R 1996 Tetrahedral amorphous carbon films prepared by magnetron sputtering and dc ion plating *J. Appl. Phys.* **79** 1416–22

[45]    Suk J W, Murali S, An J and Ruoff R S 2012 Mechanical measurements of ultra-thin amorphous carbon membranes using scanning atomic force microscopy *Carbon* **50** 2220–5

[46]    Ferrari A C, Robertson J, Beghi M G, Bottani C E, Ferulano R and Pastorelli R 1999 Elastic constants of tetrahedral amorphous carbon films by surface Brillouin scattering *Appl. Phys. Lett.* **75** 1893

[47]    Martıínez E, Andújar J L, Polo M C, Esteve J, Robertson J and Milne W I 2001 Study of the mechanical properties of tetrahedral amorphous carbon films by nanoindentation and nanowear measurements *Diam. Relat. Mater.* **10** 145–52

[48] Cho S, Chasiotis I, Friedmann T A and Sullivan J P 2005 Young's modulus, Poisson's ratio and failure properties of tetrahedral amorphous diamond-like carbon for MEMS devices *J. Micromech. Microeng.* **15** 728–735

[49] Lo R Y and Bogy D B 1999 Compensating for elastic deformation of the indenter in hardness tests of very hard materials *J. Mater. Res.* **14** 2276–82

[50] Bhushan B 1999 Chemical, mechanical and tribological characterization of ultra-thin and hard amorphous carbon coatings as thin as 3.5 nm: recent developments *Diam. Relat. Mater.* **8** 1985–2015

[51] Yin M T and Cohen M L 1981 Ground-state properties of diamond *Phys. Rev. B* **24** 6121–4

[52] Ferrari A C, Rodil S E, Robertson J and Milne W I 2002 Is stress necessary to stabilise sp3 bonding in diamond-like carbon? *Diam. Relat. Mater.* **11** 994–9

[53] Yin M T and Cohen M L 1984 Structural theory of graphite and graphitic silicon *Phys. Rev. B* **29** 6996–8

[54] Furthmüller J, Hafner J and Kresse G 1994 Ab initio calculation of the structural and electronic properties of carbon and boron nitride using ultrasoft pseudopotentials *Phys. Rev. B* **50** 15606–22

[55] Plimpton S 1995 Fast parallel algorithms for short-range molecular dynamics *J. Comput. Phys.* **117** 1–19

[56] Hutter J, Iannuzzi M, Schiffmann F and VandeVondele J 2014 cp2k: atomistic simulations of condensed matter systems *WIREs Comput. Mol. Sci.* **4** 15–25

[57]     Hjorth Larsen A, Mortensen J J, Blomqvist J, Castelli I E, Christensen R, Dułak M, Friis J, Groves M N, Hammer B, Hargus C, Hermes E D, Jennings P C, Bjerre Jensen P, Kermode J, Kitchin J R, Leonhard Kolsbjerg E, Kubal J, Kaasbjerg K, Lysgaard S, Bergmann Maronsson J, Maxson T, Olsen T, Pastewka L, Peterson A, Rostgaard C, Schiøtz J, Schütt O, Strange M, Thygesen K S, Vegge T, Vilhelmsen L, Walter M, Zeng Z and Jacobsen K W 2017 The atomic simulation environment-a Python library for working with atoms *J. Phys. Condens. Matter* **29** 273002

[58]     Lars Pastewka and James Kermode *matscipy*

[59]     Stukowski A 2010 Visualization and analysis of atomistic simulation data with OVITO--the Open Visualization Tool *Modell. Simul. Mater. Sci. Eng.* **18** 15012